\documentclass[pra,twocolumn,letterpaper,nofootinbib]{revtex4-1}
\usepackage{graphicx,amsmath,txfonts,dsfont}
\usepackage[colorlinks, linkcolor=blue, citecolor=blue, urlcolor=blue, breaklinks=true]{hyperref}

\newcommand{\Eref}[1]{Equation~(\ref{eq:#1})}
\newcommand{\eref}[1]{Eq.~(\ref{eq:#1})}
\newcommand{\fref}[1]{Fig.~\ref{fig:#1}}

\newcommand{\tref}[1]{Table~\ref{tab:#1}}

\def\rmd{\mathrm{d}}

\makeatletter
\newlength \figurewidth
\makeatother

\DeclareMathOperator{\arctanh}{arctanh}
\DeclareMathOperator{\sinc}{sinc}

\usepackage{mathtools}
\makeatletter
\newcommand{\vast}{\bBigg@{4}}
\newcommand{\Vast}{\bBigg@{5}}
\newcommand{\vastl}{\mathopen\vast}

\newcommand{\vastr}{\mathclose\vast}

\makeatother

\begin{document}

\title{``Mechano-optics'': An optomechanical quantum simulator}

\date{\today}

\author{David Edward Bruschi}
\email[Email address:\ ]{david.edward.bruschi@gmail.com}
\affiliation{York Centre for Quantum Technologies, University of York, YO10\,5DD Heslington, UK}
\affiliation{Department of Physics, University of Vienna, 1090 Vienna, Austria}
\author{Andr\'e Xuereb}
\email[Email address:\ ]{andre.xuereb@um.edu.mt}
\affiliation{Department of Physics, University of Malta, Msida MSD\,2080, Malta}

\begin{abstract}
A widely-known paradigm in optomechanical systems involves coupling the square of the position of a mechanical oscillator to an electromagnetic field. We discuss how, in the so-called resolved sideband regime, this system allows to simulate dynamics similar to ordinary optomechanics, where the position of the oscillator is coupled to the field, but with the roles of the oscillator and the field interchanged. We show that realisation of this system is within reach, and that it opens the door to an otherwise inaccessible parameter regime.
\end{abstract}

\maketitle

\section{Introduction}
The field of optomechanics\footnote{We take ``optomechanics'' here to also include systems based on microwaves and circuit QED} has in recent years achieved some long sought-after milestones~\cite{Aspelmeyer2014,Meystre2013}. Following the first observations of the cooling of a mechanical oscillator using radiation pressure~\cite{Gigan2006} came that of strong coupling~\cite{Groblacher2009a}, and a host of architectures have appeared that incorporated mechanical elements in optical or microwave cavities. Cooling of a mechanical oscillator to its ground state first by cryogenic cooling~\cite{OConnell2010} and then by means of radiation pressure~\cite{Chan2011} opened the door to experimenting with solid-state mechanical systems in the quantum regime, culminating in the observation of squeezed states of motion~\cite{Pirkkalainen2015,Lei2016} and mechanical entanglement~\cite{Riedinger2017}.

One of the long sought-after goals of optomechanics is to demonstrate manifestly quantum-mechanical behavior in the motion of a macroscopic mechanical oscillator. This has given rise to proposals discussing how, for example, one may observe jumps in the occupation number of the mechanical oscillator by monitoring the electromagnetic field leaking out of a cavity~\cite{Walls1985,Gangat2011}. The standard model within which this is explored is the so-called membrane-in-the-middle system~\cite{Thompson2008}, where a reflective membrane is placed at a node or antinode of a cavity field. All such models share a common interaction Hamiltonian that couples the photon number of the light field---$\hat{a}^\dagger\hat{a}$, with $\hat{a}$ being the corresponding annihilation operator---to the square of the position quadrature of the membrane, i.e., $\hat{x}^2$. The interaction Hamiltonian therefore reads $\hat{H}_\text{quad}=\hbar\,g\,\hat{a}^\dagger\hat{a}\,\hat{x}^2$, wiere $g$ is a, typically small, parameter that quantifies the strength of the interaction. This is the model that will form the basis of this paper. It stands in contrast to the more widely-studied linear optomechanics model~\cite{Aspelmeyer2014}, where the interaction Hamiltonian takes the form $\hat{H}_\text{lin}=\hbar\,g\,\hat{a}^\dagger\hat{a}\,\hat{x}$. The most significant limitation of this latter interaction is that $g$ is typically very small compared to the other frequency scales of the problem. This requires that one must consider the case where the cavity field has a macroscopic coherent component $\alpha$ (assumed real and positive for simplicity), such that to lowest order $\hat{H}_\text{lin}\approx\hbar\,G\,\bigl(\hat{a}+\hat{a}^\dagger\bigr)\,\hat{x}$, where $G=\alpha g\gg g$ is an amplified coupling constant. The key drawback of operating under these conditions is that the resulting $\hat{H}_\text{lin}$ is, to a very good approximation, quadratic in the operators. As a result, initially Gaussian states (which are ubiquitous in nature and which tend to be quasi-classical) remain Gaussian at all times, making it exceedingly difficult to observe non-classical behavior. This is the problem that we will tackle in this paper, by turning a quadratically-coupled optomechanical system into a quantum simulator (cf.\ also Ref.~\cite{Tacchino2017}).

\begin{figure}[b]
 \centering
  \includegraphics[width=\figurewidth]{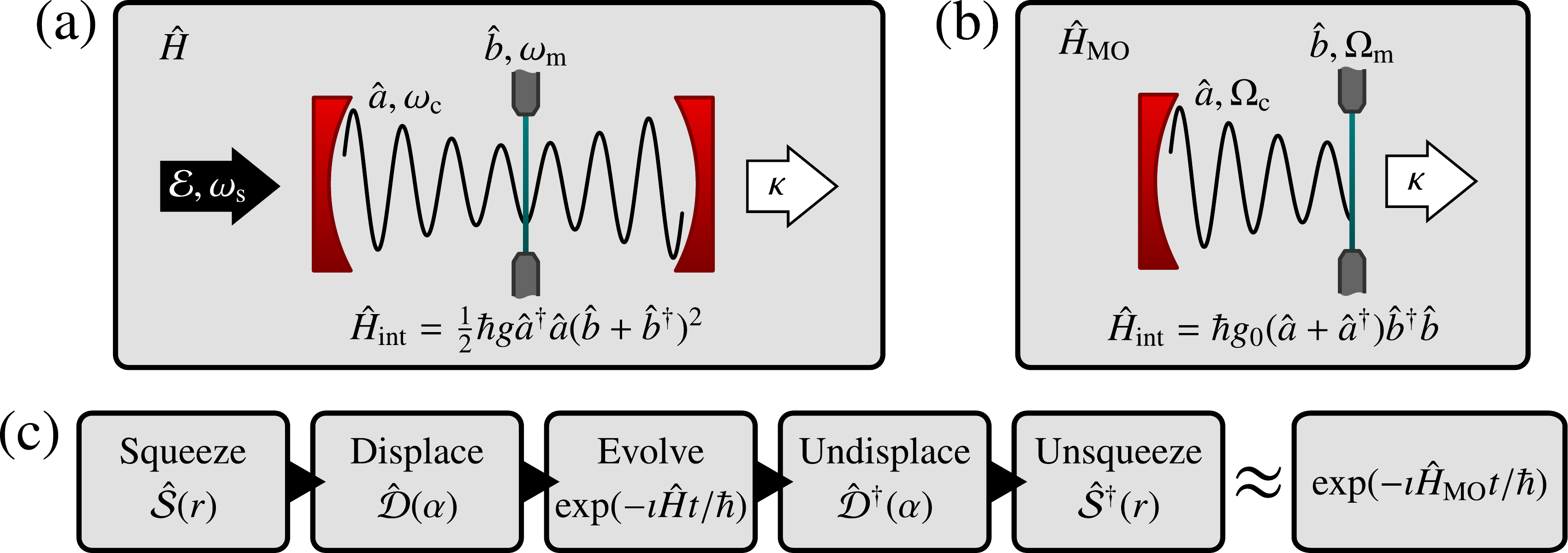}
 \caption{The system we consider in this work. (a)~The simulator takes the form of a quadratically-coupled optomechanical system, such as a membrane-in-the-middle setup~\cite{Thompson2008}. (b)~Suitable driving emulates a linearly-coupled optomechanical system with the roles of the light and mechanics reversed. (c)~The full protocol we consider requires squeezing and displacement operations to simulate the ``mechano-optical'' Hamiltonian. $\hat{H}_\text{int}$ symbolises the respective interaction Hamiltonian.}
 \label{fig:System}
\end{figure}

\section{Model}
We will consider a system consisting of one mode of the electromagnetic field coupled to one mechanical oscillator (\fref{System}). In the following, the operator $\hat{a}$ will denote the annihilation operator of the field and $\omega_\text{c}$ its frequency. The operator $\hat{x}$ ($\hat{p}$) will denote the position (momentum) of the oscillator, the parameter $m$ its mass, and $\omega_\text{m}$ its frequency. The free Hamiltonian of the system can be written $\hat{H}_\text{free}^\prime=\hbar\,\omega_\text{c}\,\hat{a}^\dagger\hat{a}+\tfrac{1}{2}\,m\,\omega_\text{m}\,\hat{x}^2+\tfrac{\hat{p}^2}{2\,m}$. Introducing the annihilation operator for the mechanical field, $\hat{b}$, through the relations $\hat{x}=\sqrt{\hbar/(2\,m\,\omega_\text{m})}\bigl(\hat{b}+\hat{b}^\dagger\bigr)$ and $\hat{p}=-\imath\sqrt{\hbar\, m\,\omega_\text{m}/2}\bigl(\hat{b}-\hat{b}^\dagger\bigr)$, allows us to write $\hat{H}_\text{free}^\prime=\hbar\,\omega_\text{c}\,\hat{a}^\dagger\hat{a}+\hbar\,\omega_\text{m}\,\bigl(\hat{b}^\dagger\hat{b}+\tfrac{1}{2}\bigr)$.

We assume that the field is driven by a classical source of strength $\mathcal{E}$ at a frequency $\omega_\text{s}$. This is modelled by means of a Hamiltonian $\hat{H}_\text{dr}^\prime=\hbar\bigl(\mathcal{E}^\ast\,e^{\imath\,\omega_\text{s}\,t}\,\hat{a}+\mathcal{E}\,e^{-\imath\,\omega_\text{s}\,t}\,\hat{a}^\dagger\bigr)$. It is convenient to transform to a frame rotating at the frequency of the source, and to neglect constant terms; we will drop the prime symbol to denote Hamiltonians in this rotating frame. This yields $\hat{H}_\text{free}=-\hbar\,\Delta\,\hat{a}^\dagger\hat{a}+\hbar\,\omega_\text{m}\,\hat{b}^\dagger\hat{b}$ and $\hat{H}_\text{dr}=\hbar\bigl(\mathcal{E}^\ast\,\hat{a}+\mathcal{E}\,\hat{a}^\dagger\bigr)$, where $\Delta=\omega_\text{s}-\omega_\text{c}$, which could be negative or positive, is the detuning between the source and field frequencies.

We also assume that the interaction term between the field and oscillator is quadratic and reads $\hat{H}_\text{quad}=\hbar\,\tilde{g}\,\hat{a}^\dagger\hat{a}\,\hat{x}^2=\hbar\,g\,\hat{a}^\dagger\hat{a}\,\bigl[\hat{b}^\dagger\hat{b}+\tfrac{1}{2}\bigl(\hat{b}^2+\hat{b}^{\dagger2}+1\bigr)\bigr]$, where the coupling strength $g$ is defined as $g:=\hbar\,\tilde{g}/(2\,m\,\omega_\text{m})$. Next, define $\hat{H}:=\hat{H}_\text{free}+\hat{H}_\text{dr}+\hat{H}_\text{quad}$, which governs the unitary evolution of the system. The cavity field is assumed to couple to the external electromagnetic vacuum at a rate $\kappa$. This is conveniently modelled using an open systems formalism whereby the system is best described by means of its density matrix $\rho$, which obeys the master equation $\dot\rho=\tfrac{1}{\imath\hbar}\bigl[\hat{H},\rho\bigr]+\mathcal{L}[\rho]$, with $\mathcal{L}[\rho]:=\kappa\,\bigl[2\,\hat{a}\,\rho\,\hat{a}^\dagger-\bigl(\hat{a}^\dagger\hat{a}\,\rho+\rho\,\hat{a}^\dagger\hat{a}\bigr)\bigr]$. A set of non-unitary terms similar to $\mathcal{L}[\rho]$ exists acting on the mechanical motion; however, as a simplifying assumption we will confine ourselves to evolution times significantly smaller than the decoherence time of the mechanical oscillator, which in turn allows us to ignore these terms in the master equation. It is useful to displace the field operators by a complex number $\alpha$. For this purpose define $\hat{\mathcal{D}}(\alpha):=\exp\bigl(\alpha\,\hat{a}^\dagger-\alpha^\ast\,\hat{a}\bigr)$, such that $\hat{\mathcal{D}}^\dagger(\alpha)\,\hat{a}\,\hat{\mathcal{D}}(\alpha)=\hat{a}+\alpha$. For reasons that will become clearer in the next step, we will introduce a squeezing operator, $\hat{\mathcal{S}}(z):=\exp\bigl[-\tfrac{1}{2}\bigl(z^\ast\hat{b}^2-z\hat{b}^{\dagger2}\bigr)\bigr]$, which acts on the mechanical state to yield, e.g., $\hat{\mathcal{S}}^\dagger(z)\,b\,\hat{\mathcal{S}}(z)=\cosh(r)\,\hat{b}+e^{\imath\theta}\sinh(r)\,\hat{b}^\dagger$, where $z=r\,e^{\imath\theta}$.

Choose $\alpha=-\mathcal{E}/\bigl(\Omega_\text{c}-\imath\kappa\bigr)$ self-consistently, where we have defined an effective frequency $\Omega_\text{c}:=-\Delta+\tfrac{1}{2}\,g\,\sqrt{\omega_\mathrm{m}\big/\bigl(\omega_\mathrm{m}+g\,\lvert\alpha\rvert^2\bigr)}$, and set $r=-\tfrac{1}{2}\,\arctanh\bigl[g\,\lvert\alpha\rvert^2\,\big/\bigl(\omega_\mathrm{m}+g\,\lvert\alpha\rvert^2\bigr)\bigr]$. Finally, define $\hat{H}_{\mathcal{DS}}:=\bigl[\hat{\mathcal{D}}(\alpha)\hat{\mathcal{S}}(r)\bigr]^\dagger\hat{H}\bigl[\hat{\mathcal{D}}(\alpha)\hat{\mathcal{S}}(r)\bigr]$. This achieves two goals: (i)~it removes a spurious mechanical squeezing term in the Hamiltonian and (ii)~it eliminates all the terms in the master equation linear in the field operators. 

By means of a suitable choice for the phase reference for $\mathcal{E}$, we can assume that $\alpha$ is real for convenience. We find that
\begin{multline}
\label{eq:HDS}
\hat{H}_{\mathcal{DS}}=\hbar\,\Omega_\text{c}\,\hat{a}^\dagger\hat{a}+\hbar\,\Omega_\text{m}\,\hat{b}^\dagger\hat{b}+\hbar g_0\,\bigl(\hat{a}+\hat{a}^\dagger\bigr)\hat{b}^\dagger\hat{b}\\
+\hbar\,g\,\hat{a}^\dagger\hat{a}\,\bigl[\hat{b}^\dagger\hat{b}+\tfrac{1}{2}\,\bigl(\hat{b}^2+\hat{b}^{\dagger2}\bigr)\bigr]+\tfrac{\hbar}{2}\,g_0\,\bigl(\hat{a}+\hat{a}^\dagger\bigr)\bigl(\hat{b}^2+\hat{b}^{\dagger2}\bigr),
\end{multline}
where $g_0:=g\,\alpha$ and $\Omega_\text{m}:=\sqrt{\omega_\text{m}^2+2\,g\,\omega_\text{m}\,\lvert\alpha\rvert^2}$, and where we omitted constant terms. 

We now argue that the terms in the second line of the above equation can be ignored. First, we will consider the situation where $g$ is very small, such that a large $\lvert\alpha\rvert$ is used in order to effectively amplify the interaction; this is the situation most frequently encountered in present-day optomechanical systems~\cite{Aspelmeyer2014}. Terms of order $\lvert g\rvert=\lvert g_0/\alpha\rvert\ll\lvert g_0\rvert$ can therefore be ignored safely. We also assume that $\lvert\alpha\rvert$ is, conversely, small enough to avoid any bistable or unstable dynamics (cf.\ Applications, below). Second, because we will also assume that $\Omega_\mathrm{m}\gg\kappa\sim\Omega_\mathrm{c}$ it is possible to invoke the rotating-wave approximation (RWA)~\cite{WilsonRae2007,Marquardt2007}, also commonly used in optomechanics, to drop the second term. We note at this point that applying the rotating-wave approximation directly to $\hat{H}_\text{quad}$ neglects the mechanical squeezing produced by the mean field $\alpha$, and will therefore give incorrect results when $\lvert\alpha\rvert$ is large.

In the next section we will show that the evolution operator derived from $\hat{H}_\mathcal{DS}$ is approximately equal to one derived from a greatly simplified model corresponding to the usual optomechanical Hamiltonian, but with the optical and mechanical fields interchanged.

\begin{table*}[ht!]
\begin{ruledtabular}
\begin{tabular}{lcccccc}
 & Oscillation & Mechanical & Base tem- & Cavity field & Mechanical & Quadratic opto-\\
 & frequency & decay rate & perature & HWHM linewidth & squeezing (max.) & mechanical coupling \\
 Parameter & $\omega_\text{m}$ & $\gamma_\text{m}$ & $T$ & $\kappa$ & $r_\text{max}$ & $g$\\
 \hline
 Mechanics & $2\pi\times140$\,kHz & $2\pi\times1.4$\,mHz & $500$\,mK & $2\pi\times70$\,kHz & $-0.54$ & $5.2\times10^{-4}/$s \\
 Reference & \cite{Norte2016} & \cite{Norte2016} & \cite{Norte2016} & \cite{Karuza2012,Lee2015} & \cite{Lei2016} & \cite{Norte2016,Lee2015}\\
 \hline
 cQED & $2\pi\times300$\,MHz & $2\pi\times17$\,kHz & $10$\,mK & $2\pi\times330$\,kHz & $-0.54$ & $19\times10^3/$s \\
 Reference & \cite{Kim2015} & \cite{Lei2016} & \cite{Lei2016} & \cite{Lei2016} & \cite{Lei2016} & \cite{Kim2015}
\end{tabular}
\end{ruledtabular}
\caption{The numerical values used to illustrate the feasibility of the system presented here. Two sets of numbers are shown; the first (``Mechanics'') refers to an optomechanical system with a macroscopic mechanical oscillator in an optical cavity, whereas the second (``cQED'') is based on an electromechanical system that reproduces the quadratic optomechanical Hamiltonian, albeit with coupling strengths orders of magnitude larger than in the optical domain.}
\label{tab:Numbers}
\end{table*}

\emph{From $\hat{H}_\mathcal{DS}$ to ``mechano-optics.''}---We want to compute the time evolution operator $\hat{U}_\mathcal{DS}(t):=\exp(-\imath\hat{H}_\mathcal{DS}t/\hbar)$ induced by the Hamiltonian \eqref{eq:HDS}. Operating under the assumption that $g/\Omega_\text{m}\ll1$, and that $\lvert\alpha\rvert\gg1$, we can treat the term $\hat{H}_\text{small}:=\hbar\,g\,\hat{a}^\dagger\hat{a}\,\bigl[\hat{b}^\dagger\hat{b}+\tfrac{1}{2}\,\bigl(\hat{b}^2+\hat{b}^{\dagger2}\bigr)\bigr]$ as a small perturbation. Next, we define the ``mechano-optical'' Hamiltonian
\begin{equation}
\hat{H}_\text{MO}:=\hbar\Omega_\text{c}\hat{a}^\dagger\hat{a}+\hbar\Omega_\text{m}\hat{b}^\dagger\hat{b}+\hbar g_0\bigl(\hat{a}^\dagger+\hat{a}\bigr)\hat{b}^\dagger\hat{b},
\end{equation}
and the auxiliary term $\hat{H}_\text{aux}:=\tfrac{\hbar}{2}g_0\bigl(\hat{a}+\hat{a}^\dagger\bigr)\bigl(\hat{b}^2+\hat{b}^{\dagger2}\bigr)$, such that $\hat{H}_\mathcal{DS}=\hat{H}_\text{MO}+\hat{H}_\text{aux}+\hat{H}_\text{small}$. Operating under the assumption that $\hat{H}_\text{small}$ is a perturbation, it is possible to write a concise expression for $\hat{U}_\mathcal{DS}(t)$ in terms of the evolution operator $\hat{U}_\text{MO}(t):=\exp(-\imath\hat{H}_\text{MO}t/\hbar)$; this calculation is detailed elsewhere~(see Appendix). Our immediate aim is to quantify how similar the evolution of a state under the action of $\hat{U}_\text{MO}(t)$ is to that under $\hat{U}_\mathcal{DS}(t)$. Let us introduce the fidelity $\mathcal{F}(t):=\lvert\langle\psi_\mathcal{DS}(t)\vert\psi_\text{MO}(t)\rangle\rvert^2$, which measures the overlap between the states $\lvert\psi_\mathcal{DS}(t)\rangle=\hat{U}_\mathcal{DS}(t)\lvert\psi_0\rangle$ and $\lvert\psi_\text{MO}(t)\rangle=\hat{U}_\text{MO}(t)\lvert\psi_0\rangle$ for some arbitrary initial pure state $\lvert\psi_0\rangle$. We find, after some calculations~(see Appendix), that
\begin{align}
\label{eq:FidelityPerturbative}
\mathcal{F}(t)=1+\bigl[\langle\psi_0\vert\hat{E}_1(t)\vert\psi_0\rangle^2-\langle\psi_0\vert\hat{E}^2_1(t)\vert\psi_0\rangle\bigr],
\end{align}
to second order in the small parameter $g_0/\Omega_\text{m}$, where we have introduced
\begin{multline}
\hat{E}_1(t)=\tfrac{1}{2}\int_0^t\rmd t^\prime\,g_0(t^\prime)\bigl(\hat{a}\,e^{-\imath\,\Omega_\text{c}\,t^\prime}+\hat{a}^\dagger\,e^{\imath\,\Omega_\text{c}\,t^\prime}\bigr)\\
\times\bigl(\hat{b}^2\,e^{-2\,\imath\,\Omega_\text{m}\,t^\prime}+\hat{b}^{\dagger2}\,e^{2\,\imath\,\Omega_\text{m}\,t^\prime}\bigr),
\end{multline}
allowing $g_0$ to inherit an explicit time-dependence from $\alpha$. \Eref{FidelityPerturbative} may be simplified further for times $t\ll1/\max_t\{\lvert g_0\rvert\}$, yielding
\begin{equation}
\mathcal{F}\bigl(t\ll\tfrac{1}{\max_t\{\lvert g_0\rvert\}}\bigr)=1-F_\text{uni}(t)-\langle\psi_0\vert\hat{E}_\text{NO}(t)\vert\psi_0\rangle,
\end{equation}
where $\hat{E}_\text{NO}(t)$ is a normally-ordered sixth-order polynomial function of $\hat{a}$, $\hat{a}^\dagger$, $\hat{b}$, and $\hat{b}^\dagger$ that is of order $\bigl(g_0/\Omega_\text{m}\bigr)^2$. The function $F_\text{uni}(t)\geq0$ is a universal quantity independent of the initial state. For constant $g_0$ it can be written as
\begin{align}
F_\text{uni}(t)=2\,g_0^2\,\Biggl[\frac{\sin^2\bigl(\Omega_+t\bigr)}{\Omega^2_+}+\frac{\sin^4\bigl(\Omega_-t/2\bigr)}{\Omega^2_-/4}\Biggr],
\end{align}
with $\Omega_{\pm}=\Omega_\text{c}\pm2\,\Omega_\text{m}$. Starting from the vacuum state $\lvert\psi_0\rangle=\lvert0\rangle$ we therefore obtain
\begin{equation}\label{vacuum:fidelity}
\mathcal{F}_\text{vac}\bigl(t\ll\tfrac{1}{\max_t\{\lvert g_0\rvert\}}\bigr)=1-F_\text{uni}(t).
\end{equation}
Our result \eqref{vacuum:fidelity} allows us, from now on, to consider situations where $\mathcal{F}(t)\approx1$ during the relevant time period. Therefore, to a good approximation, we can claim that the Hamiltonian governing the system is $\hat{H}_\text{MO}$, which is identical in form to the usual linear optomechanical interaction Hamiltonian with $g_0$ playing the role of the single-photon coupling rate and---crucially---the roles of the optical field and mechanical oscillator reversed. This is the central result of this paper and yields what we will refer to as ``mechano-optical'' dynamics. Explicity, $\hat{H}_{\mathcal{DS}}\approx\hat{H}_\text{MO}$. Note also that the three frequencies $\Omega_\text{m}$, $\Omega_\text{c}$, and $g_0$ are all independent free parameters of the model. From this point on, we shall assume that $g_0$ is constant.

Since $\hat{\mathcal{S}}(r)$ and $\hat{\mathcal{D}}(\alpha)$ are unitary operations it follows that, as illustrated in \fref{System},
\begin{equation}
\hat{\mathcal{S}}^\dagger(r)\,\hat{\mathcal{D}}^\dagger(\alpha)\,\exp\bigl(-\imath\hat{H}t/\hbar\bigr)\,\hat{\mathcal{D}}(\alpha)\,\hat{\mathcal{S}}(r)\approx\exp\bigl(-\imath\hat{H}_\text{MO}t/\hbar\bigr).
\end{equation}
At this point we note two things about the effective Hamiltonian $\hat{H}_\text{MO}$. First, the sign and magnitude of the single-\emph{phonon} coupling strength, $g_0$, can now be set at will by means of an appropriate choice of the driving strength $\mathcal{E}$. This was noted, but not discussed at length, in the context of a somewhat related model studied in Ref.~\cite{Zhang2012}. This freedom of choice allows us to implement experimental protocols that cannot be performed otherwise. For example, it allows us to simulate the non-equilibrium thermodynamics of optomechanical systems undergoing sudden quenches~\cite{Fusco2014} by turning this interaction on or off as required. Second, the effective single-photon coupling strength in this model may be made very large. The system presented here therefore provides a much-needed short-cut towards simulating strong-coupling physics and makes it possible to effectively enter the single-photon strong-coupling regime of optomechanics; something which, despite significant theoretical~\cite{Ludwig2012,Xuereb2012c,Xuereb2013} and experimental~\cite{Massel2012,Pirkkalainen2014,Yuan2015b} progress in recent years, has thus far proven to be elusive, although atom-optomechanical systems have started approaching this regime~\cite{Brooks2012}.

\section{Realisability}
The key approximations made in the above derivation were two. First, we assumed that conditions are such that the terms in the second line of \eref{HDS} can be ignored. Our fidelity calculations provide justification for the correctness of this statement. Second, we assumed that the entire protocol and evolution can be performed in a time much shorter than the mechanical decoherence time, such that no mechanical dissipative terms need to be included in the master equation. Ref.~\cite{Lei2016} demonstrates a mechanical oscillator with frequency $\omega_\text{m}=2\pi\times5.8$\,MHz and mechanical linewidth $\gamma_\text{m}=2\pi\times8$\,Hz operating in a dilution refrigerator at a base temperature $T=10$\,mK, the number of phonons in steady-state is $n_\text{p}=1/\bigl[e^{\hbar\omega_\text{m}/(k_\text{B}T)}-1\bigr]\approx35$ ($k_\text{B}$ is Boltzmann's constant). This yields a decoherence time of $1/(\gamma_\text{m}n_\text{p})\approx0.6$\,ms, which equates to $n_\text{osc}\gtrsim20\,000$ oscillation periods. The greatest squeezing generated in Ref.~\cite{Lei2016} is reported as $4.7\pm0.9$\,dB; the mean corresponds to a maximal squeezing parameter $\lvert r_\text{max}\rvert=-r_\text{max}=-\tfrac{1}{2}\ln\bigl(10^{-4.7/10}\bigr)\approx0.54$.

A mechanical oscillator closer in form to the one illustrated in \fref{System} is explored in Ref.~\cite{Norte2016}, which has a motional mass $m=1$\,ng, $\omega_\text{m}=2\pi\times140$\,kHz, $\gamma_\text{m}=2\pi\times1.4$\,mHz and at $T=500$\,mK~\cite{Lee2015} ($n_\text{p}\approx74\,000$) has a decoherence time equating to $n_\text{osc}\gtrsim1\,300$ oscillations. Temperatures down to $14$\,mK ($n_\text{osc}\gtrsim47\,000$) are achievable~\cite{Yuan2015}, but uncommon, for this kind of system. Using the second derivative of the cavity mode frequency from Ref.~\cite{Lee2015} and the mechanical parameters from Ref.~\cite{Norte2016} yields $g=5.2\times10^{-4}/$s for an effective quadratic coupling generated by exploiting avoided crossings between cavity resonances. The cavity is assumed to have a finesse of $60\,000$, consistent with the values reported in Ref.~\cite{Karuza2012} for a similar setup, yielding a half-width at half-maximum linewidth $\kappa\approx2\pi\times70\text{\,kHz}\ll\omega_\text{m}$.

\begin{figure*}[t]
\centering
 \includegraphics[width=0.2\textwidth]{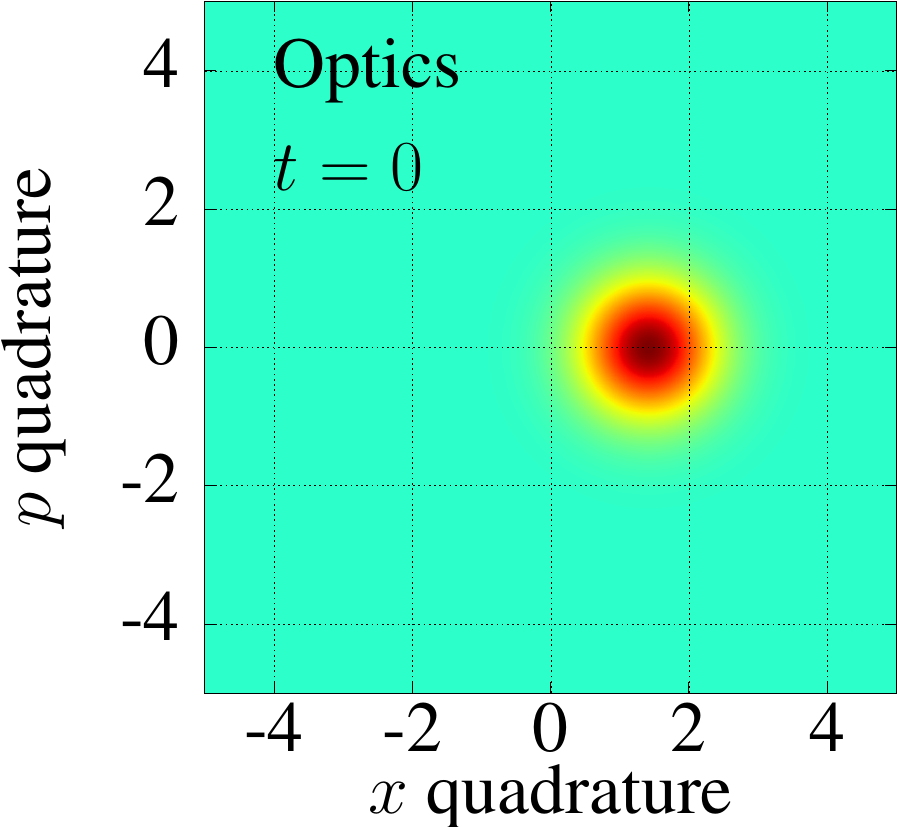}\qquad
 \includegraphics[width=0.2\textwidth]{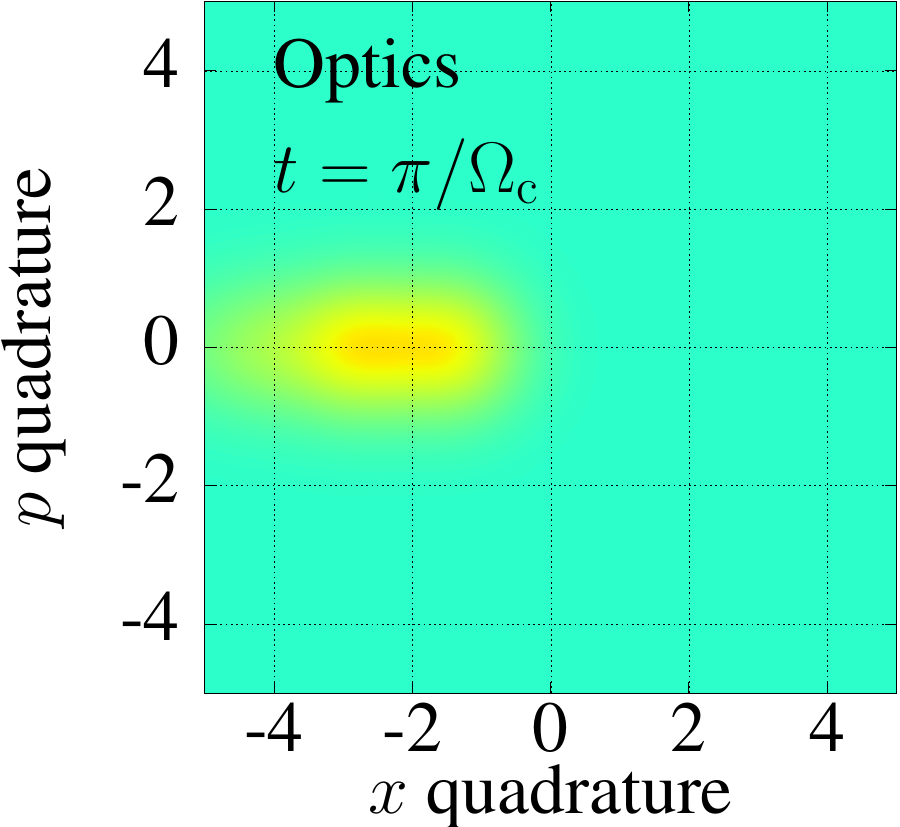}\qquad
 \includegraphics[width=0.2\textwidth]{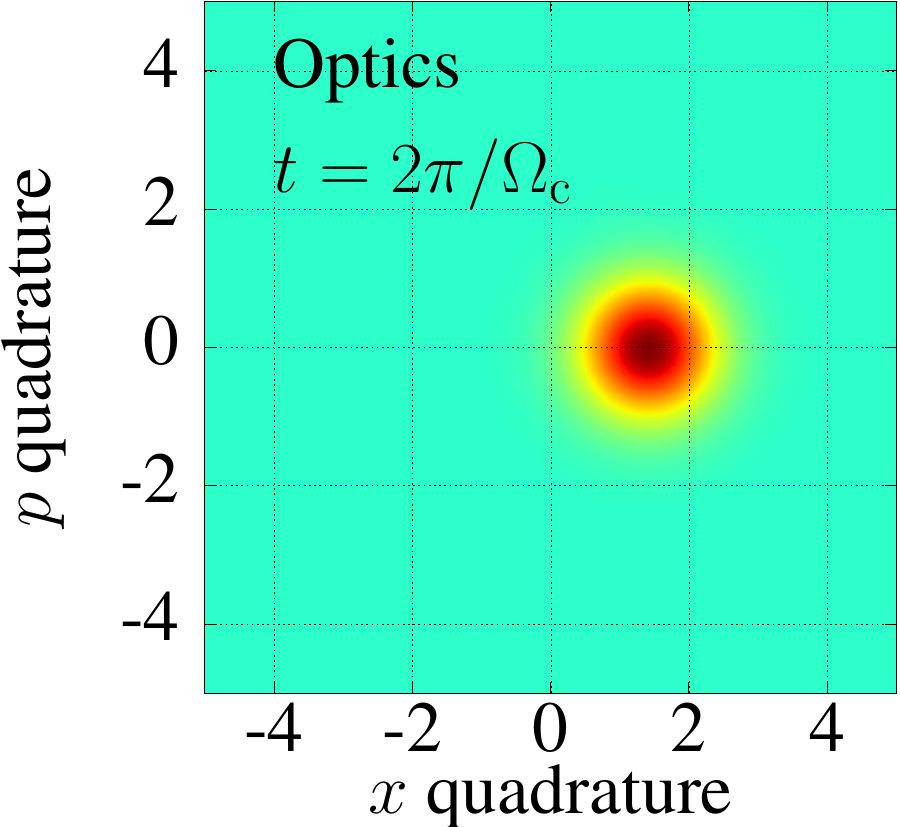}\\[2mm]
 \includegraphics[width=0.2\textwidth]{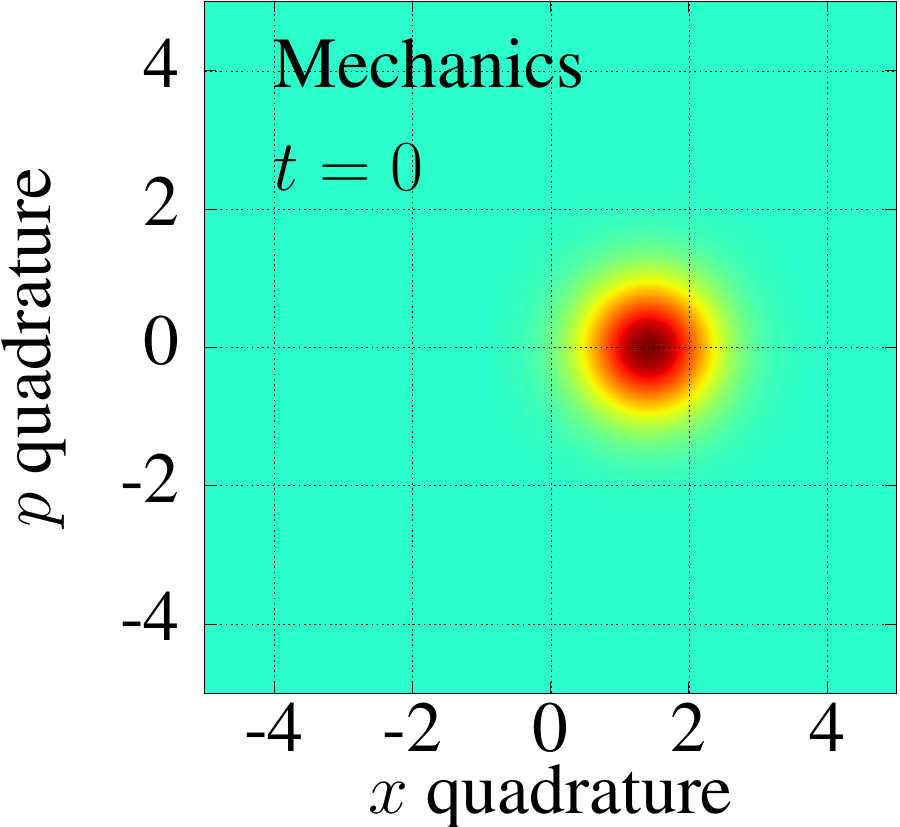}\qquad
 \includegraphics[width=0.2\textwidth]{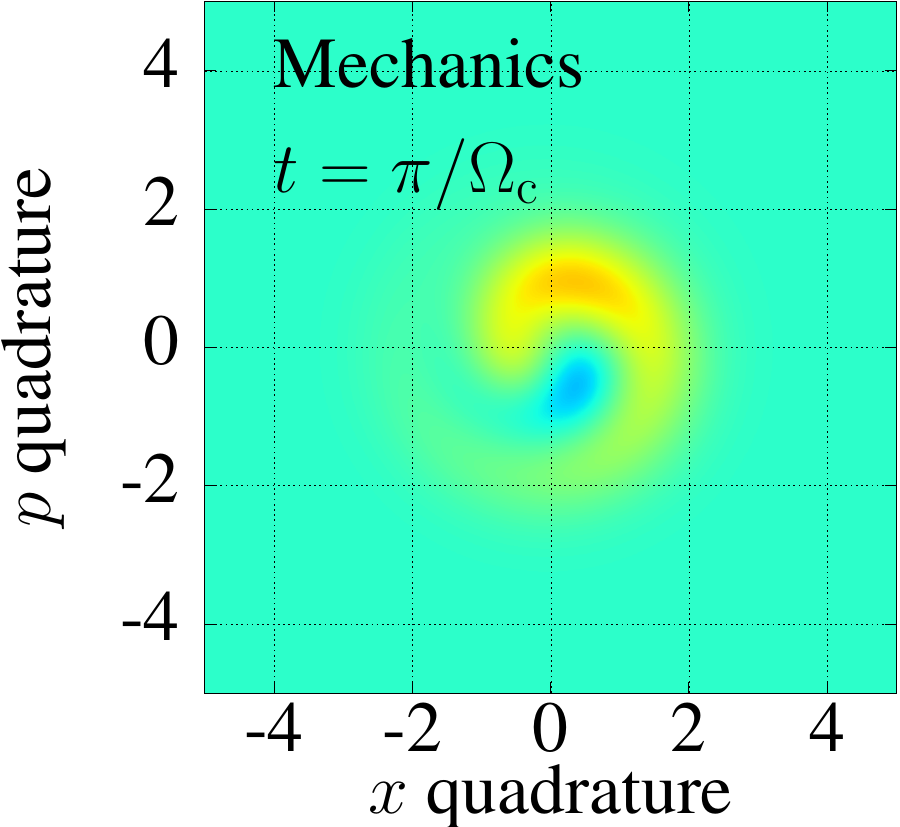}\qquad
 \includegraphics[width=0.2\textwidth]{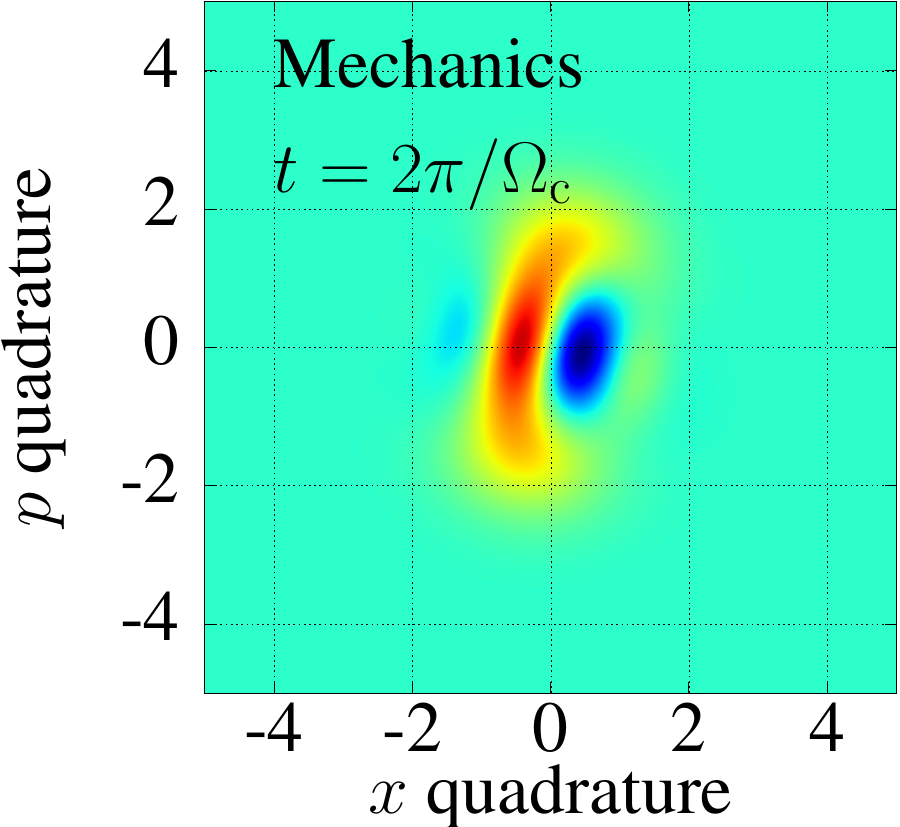}
\caption{(Color online.) Simulating the single-photon strong-coupling regime of optomechanics. The first (second) row represents the Wigner function of the optical (mechanical) field; from left to right the images show $t=0$, $\pi/\Omega_\text{c}$, and $2\pi/\Omega_\text{c}$. Here we show the ``cQED'' system (cf.\ \tref{Numbers}) explicitly but the situation is very similar for the ``Mechanics'' system, with appropriately chosen parameters. The right-most figures should be compared with Figs.~2(e) and~3(a) in Ref.~\cite{Bose1997}, respectively; note that the optical and mechanical fields have had their roles switched. The parameters have been chosen such that $g_0=0.5\Omega_\text{c}$, and the initial state as product of coherent states $\lvert1\rangle_\text{a}\otimes\lvert1\rangle_\text{b}$.}
\label{fig:Simulation}
\end{figure*}

Current mechanical realisations of quadratically-coupled systems suffer from a very small quadratic coupling strength normalised to the cavity linewidth, $g/\kappa$. A means for overcoming this problem was suggested in Ref.~\cite{Kim2015}, using cavity QED (``cQED'') techniques to realise an all-electronic analog of a quadratically-coupled optomechanical system. This allows to achieve relative coupling strengths many orders of magnitude larger than would otherwise be possible. Combining this technique with experimentally-achieved numbers from Ref.~\cite{Lei2016}, it is possible to envisage a system with an effective mechanical frequency $\omega_\text{m}=2\pi\times300$\,MHz. Assuming a finesse of around $18\,000$~\cite{Lei2016} for both resonators yields an optical linewidth $\kappa=2\pi\times330$\,kHz and a mechanical decay rate $\gamma_\text{m}=2\pi\times17$\,kHz. Finally, using the data in Fig.~12 of Ref.~\cite{Kim2015}, with $m=2$, $n=0,1$, and $\Phi_\text{ext}^0/\Phi_0=0.4$, yields $g\approx19\times10^3$/s for this system. Operating at a base temperature of $10$\,mK yields a thermal population $n_\text{p}\approx0.3$, which implies that the effective mechanical oscillator can be assumed to be in its ground state, and $n_\text{osc}\gtrsim50\,000$, so that any decay and decoherence processes can be safely neglected. These values, summarised in \tref{Numbers}, were used for the examples that will be presented below.

\section{Applications}
In this section we will outline two applications of our techniques, which have far-reaching consequences. First, we can apply the system we discussed to simulating the dynamics of optomechanical systems in the single-photon strong-coupling regime. For concreteness, we discuss explicitly the ``cQED'' system in \tref{Numbers}; the ``Mechanics'' system can be treated similarly but requires stronger squeezing to overcome the decay of the optical field. Setting $\Omega_\text{c}=2g_0$ to approach the single-photon strong-coupling regime and choosing our parameters self-consistently ($\alpha\simeq80\,752$) we can reproduce dynamics reminiscent of the study in Ref.~\cite{Bose1997}; see \fref{Simulation} for further details. Of interest are two facts. First, despite the rather small bare coupling coefficient $g$ we have obtained effective strong-coupling dynamics. Second, the mechanical state acquires a strongly non-classical character, as shown by the strongly negative Wigner function. This is different from the standard single-photon strong-coupling regime in optomechanics, where under simple driving it is the optical field that acquires a non-classical character. We neglect decay processes in the figure because the last time-step shown is at $2\pi/\Omega_\text{c}$, which is smaller than the decay time $1/\kappa$. In the example illustrated in \fref{Simulation} the fidelity between the dynamics generated by $\hat{H}_\text{MO}$ and $\hat{H}_\mathcal{DS}$ is $\geq99.58$\%.

The second application we propose is aimed at probing the instabilities that naturally arise in this system. It is not difficult to show that the eigenvalues $\lambda_{n,l}$ of the mechano-optical Hamiltonian $\hat{H}_\text{MO}$ have the form
\begin{equation}
\lambda_{n,l}=\hbar\,\biggl[n\,\Omega_\text{c}+l\,\Omega_\text{m}\biggl(1-\tfrac{l\,g_0^2}{\Omega_\text{m}\,\Omega_\text{c}}\biggr)\biggr],
\end{equation}
where $n,l\in\mathbb{N}$, and the corresponding eigenstates $\lvert\lambda_{n,l}\rangle$ have the form $\lvert\lambda_{n,l}\rangle:=\hat{D}^\dag\bigl(l\frac{g_0}{\Omega_\text{c}}\bigr)\,\lvert n\rangle_\text{a}\otimes\lvert l\rangle_\text{b}$. For all $l\geq l_\text{max}:=\Omega_\text{m}\Omega_\text{c}/g_0^2$, these eigenvalues are negative for some values of $n$, which introduces instabilities if the state of the oscillator has significant overlap with $\lvert l\geq l_\text{max}\rangle_\text{b}$. Our scheme allows to access these regimes, which are otherwise inaccessible---using the parameters in Ref.~\cite{Lei2016}, for example, the instability is expected to become noticeable at temperatures around $0.6$\,GK, since $l_\text{max}\sim10^{12}$. In our case, however, one can bring $l_\text{max}$ to a low value and trigger instabilities in the system since it no longer has a well-defined Gibbs state. We note in passing that for large enough displacements any realisable mechanical potential becomes anharmonic; this instability is therefore not easily observable otherwise. Indeed, adding a non-linearity of the form $\hbar\chi(\hat{b}^\dagger\hat{b})^2$ to $\hat{H}_\text{MO}$, with $\chi\geq g_0^2/\Omega_\text{c}$, restores the positivity of all eigenvalues.

\section{Conclusion}
We have introduced a mechanical quantum simulator based on a quadratically-coupled optomechanical system. This system can effectively reproduce the dynamics of a standard optomechanical system, but where the roles of the optical and mechanical fields are switched, and where the single-photon coupling strength can be chosen by driving the system appropriately. As an example, we have shown how to apply our system to simulate the single-photon strong-coupling regime, and have discussed its role in exploring instabilities in quantum systems. Our work opens the door to an entirely new use case for optomechanical systems.

\section{Acknowledgment}
We thank Ch.K.\ Wick for useful comments and discussions. This work was partially supported by COST Actions MP1403, MP1405, CA15117, and CA15220. We acknowledge funding from the European Union's Horizon 2020 research and innovation program under grant agreement No.\ 732894 (FETPRO HOT). D.E.B.\ acknowledges the hospitality of the University of Malta and the University of Vienna.

\appendix\onecolumngrid\newpage
\section{Evolution operator for a time-dependent coupling strength}
We start with the Hamitonian
\begin{equation}
\hat{H}_\text{MO}(t)=\hbar\Omega_\text{c}\hat{a}^\dagger\hat{a}+\hbar\Omega_\text{m}\hat{b}^\dagger\hat{b}+\hbar g_0\bigl(\hat{a}^\dagger+\hat{a}\bigr)\hat{b}^\dagger\hat{b},
\end{equation}
where we allow $g_0$ to be time dependent, i.e., $g_0\rightarrow g_0(t)$. Define the four hermitian operators
\begin{align}
\hat{N}_a&:=\hat{a}^\dag\hat{a},\\
\hat{N}_b&:=\hat{b}^\dag\hat{b},\\
\hat{G}_+&:=(\hat{a}^\dag+\hat{a})\hat{b}^\dag\hat{b},\ \text{and}\\
\hat{G}_-&:=\imath(\hat{a}^\dag-\hat{a})\hat{b}^\dag\hat{b}.
\end{align}
The set of operators $\{\hat{N}_a,\hat{N}_b,\hat{N}_b^2,\hat{G}_+,\hat{G}_-\}$ forms a closed Lie algebra. To exploit this fact, define also
\begin{align}
\hat{U}_a(t)&:=e^{-\imath F_a(t)\hat{N}_a},\\
\hat{U}_b(t)&:=e^{-\imath F_b(t)\hat{N}_b},\\
\hat{U}^{(2)}_b(t)&:=e^{-\imath F^{(2)}_b(t)\hat{N}_b^2},\\
\hat{U}_+(t)&:=e^{-\imath F_+(t)\hat{G}_+},\ \text{and}\\
\hat{U}_-(t)&:=e^{-\imath F_-(t)\hat{G}_-},
\end{align}
where the real time-dependent functions $F_b(t)$, $F^{(2)}_b(t)$, $F_a(t)$, $F_+(t)$, and $F_-(t)$ are still be determined. It is not difficult to show that
\begin{align}
\hat{U}_a(t)\,\hat{G}_+\,\hat{U}_a^\dag(t)&=\cos[F_a(t)]\,\hat{G}_+-\sin[F_a(t)]\,\hat{G}_-,\\
\hat{U}_a(t)\,\hat{G}_-\,\hat{U}_a^\dag(t)&=\cos[F_a(t)]\,\hat{G}_-+\sin[F_a(t)]\,\hat{G}_+,\ \text{and}\\
\hat{U}_+(t)\,\hat{G}_-\,\hat{U}_+^\dag(t)&=\hat{G}_-+2F_+(t)\hat{N}_b^2.
\end{align}
Next, the evolution operator $U(t)$ corresponding to $\hat{H}_\text{MO}(t)$ is defined as
\begin{equation}
U(t)=\overset{\leftarrow}{\mathcal{T}}\exp\biggl[-\frac{\imath}{\hbar}\int^{t}_0\rmd t^\prime\,\hat{H}_\text{MO}(t^\prime)\biggr],
\end{equation}
where $\overset{\leftarrow}{\mathcal{T}}$ is the time ordering operator. Using the techniques in Ref.~\cite{Bruschi2013} we can always write this as
\begin{equation}\label{eq:Appendix:EvolutionDecomposition}
\hat{U}(t)=\hat{U}_b(t)\,\hat{U}^{(2)}_b(t)\,\hat{U}_a(t)\,\hat{U}_+(t)\,\hat{U}_-(t), 
\end{equation}
which is accompanied by the differential equation
\begin{equation}
\hat{H}_\text{MO}(t)=\dot{F}_b(t)\hat{N}_b+\dot{F}^{(2)}_b(t)\,\hat{N}_b^2+\dot{F}_a(t)\,\hat{N}_a+\dot{F}_+(t)\,\hat{U}_a(t)\,G_+\,U^\dag_a(t)+\dot{F}_-(t)\,\hat{N}_b+\dot{F}_-(t)\,\hat{U}_a(t)\,\hat{U}_+(t)\,\hat{G}_-\,\hat{U}_+^\dag(t)\,\hat{U}^\dag_a(t).
\end{equation}
It is this differential equation that determines the functions $F_b(t)$, $F^{(2)}_b(t)$, $F_a(t)$, $F_+(t)$, and $F_-(t)$, together with $F_b(0)=F^{(2)}_b(0)=F_a(0)=F_+(0)=F_-(0)=0$. Indeed, after some algebra we obtain
\begin{align}
F_a(t)&:=\Omega_\text{c}\,t,\\
F_b(t)&:=\Omega_\text{m}\,t,\\
F_b^{(2)}(t)&:=-2\,\frac{g(t)}{\sqrt{\Omega_\text{c}\Omega_\text{m}}}\,\sin(\Omega_\text{c}t^\prime)\,\int_0^t dt^\prime\,g(t^\prime)\,\cos(\Omega_\text{c}t^\prime),\\
F_+(t)&:=\int_0^t dt^\prime\,g(t^\prime)\,\cos(\Omega_\text{c}t^\prime),\ \text{and}\\
F_-(t)&:=\int_0^t dt^\prime\,g(t^\prime)\,\sin(\Omega_\text{c}t^\prime).
\end{align}
These expressions, together with the decomposition of the time evolution operator \eref{Appendix:EvolutionDecomposition}, represent a compact solution for the time evolution of a mechano-optical (or an optomechanical) system whose coupling constant is allowed to depend on time.

\section{Simplification of the time evolution operator}
We want to compute, and obtain a simplified form for, the time evolution operator
\begin{equation}
\hat{U}_\mathcal{DS}(t)=\overset{\leftarrow}{\mathcal{T}}\exp\biggl[-\frac{\imath}{\hbar}\int_0^t dt^\prime \hat{H}_\mathcal{DS}(t^\prime)\biggr],
\end{equation}
induced by the Hamiltonian
\begin{equation}
\label{eq:HDS:appendix}
\hat{H}_{\mathcal{DS}}=\hbar\Omega_\text{c}\hat{a}^\dagger\hat{a}+\hbar\Omega_\text{m}\hat{b}^\dagger\hat{b}+\hbar g_0\bigl(\hat{a}+\hat{a}^\dagger\bigr)\hat{b}^\dagger\hat{b}+\hbar g\hat{a}^\dagger\hat{a}\bigl[\hat{b}^\dagger\hat{b}+\tfrac{1}{2}\bigl(\hat{b}^2+\hat{b}^{\dagger2}\bigr)\bigr]+\tfrac{\hbar}{2}g_0\bigl(\hat{a}+\hat{a}^\dagger\bigr)\bigl(\hat{b}^2+\hat{b}^{\dagger2}\bigr).
\end{equation}
Here, the time-ordering operator $\overset{\leftarrow}{\mathcal{T}}$ is defined such that time is ordered in decreasing order from left to right. As in the main text, we will assume that the condition $\alpha\gg1$ holds. This allows us to write
\begin{equation}
\hat{U}_\mathcal{DS}(t)=\hat{U}_\text{MO}(t)\,\overset{\leftarrow}{\mathcal{T}}\exp\biggl[-\frac{\imath}{\hbar}\int_0^t \rmd t^\prime U^\dag_\text{MO}(t^\prime)\,\hat{H}_\text{aux}\,\hat{U}_\text{MO}(t^\prime)\biggr]\biggl[\mathds{1}-\frac{\imath}{\hbar}\int_0^t\rmd t^{\prime\prime} U^\dag_\text{MO}(t^{\prime\prime})\,\hat{H}_\text{small}\,\hat{U}_\text{MO}(t^{\prime\prime})\biggr]+\mathcal{O}\,\biggl(\frac{1}{\alpha^2}\biggr),
\end{equation}
where we have introduced
\begin{align}
\hat{H}_\text{small}&=\hbar g\hat{a}^\dagger\hat{a}\bigl[\hat{b}^\dagger\hat{b}+\tfrac{1}{2}\bigl(\hat{b}^2+\hat{b}^{\dagger2}\bigr)\bigr],\\
\hat{H}_\text{MO}&=\hbar\Omega_\text{c}\hat{a}^\dagger\hat{a}+\hbar\Omega_\text{m}\hat{b}^\dagger\hat{b}+\hbar g_0\bigl(\hat{a}+\hat{a}^\dagger\bigr)\hat{b}^\dagger\hat{b},\ \text{and}\\
\hat{H}_\text{aux}&=\tfrac{\hbar}{2}g_0\bigl(\hat{a}+\hat{a}^\dagger\bigr)\bigl(\hat{b}^2+\hat{b}^{\dagger2}\bigr),
\end{align}
as per the main text, and defined
\begin{equation}
\hat{U}_\text{MO}(t)=\exp\biggl(-\frac{\imath}{\hbar}\int_0^t\rmd t^\prime\hat{H}_\text{MO}(t^\prime)\biggr).
\end{equation}
To zeroth order in $1/\alpha$, we therefore have
\begin{align}
\hat{U}_\mathcal{DS}(t)=\hat{U}_\text{MO}(t)\,\overset{\leftarrow}{\mathcal{T}}\exp\biggl[-\frac{\imath}{\hbar}\int_0^t \rmd t^\prime U^\dag_\text{MO}(t^\prime)\,\hat{H}_\text{aux}\,\hat{U}_\text{MO}(t^\prime)\biggr].
\end{align}
Calculations yield
\begin{multline}
\label{eq:Appendix:UDS}
\hat{U}_\mathcal{DS}(t)=\hat{U}_\text{MO}(t)\,\overset{\leftarrow}{\mathcal{T}}\exp\biggl(-\imath\int_0^t \rmd t^\prime g_0(t^\prime)\bigl[\hat{A}_+\,\cos(\Omega_\text{c}\,t^\prime)+\hat{A}_-\,\sin(\Omega_\text{c}\,t^\prime)\bigr]\\
\times\bigl\{\hat{B}_+\,\cos\bigl[2(\Omega_\text{m}+g_0\,\hat{A}_+)t^\prime\bigr]+\hat{B}_-\,\sin\bigl[2(\Omega_\text{m}+g_0\,\hat{A}_+)t^\prime\bigr]\bigr\}\biggr),
\end{multline}
where we have allowed $g_0$ to have an explicit time-dependence and defined
\begin{align}
\hat{A}_+&:=\hat{a}^\dagger+\hat{a},\\
\hat{A}_-&:=-\imath(\hat{a}-\hat{a}^\dagger),\\
\hat{B}_+&:=\hat{b}^{\dagger2}+\hat{b}^2,\ \text{and}\\
\hat{B}_-&:=-\imath(\hat{b}^2-\hat{b}^{\dagger2}).
\end{align}
At this stage we note that our main decoupling result, \eref{Appendix:UDS}, allows us to write the time evolution operator in such a way that $\hat{U}_\mathcal{DS}(t)$ factors out. The rest of the expression can be treated as a correction term. Our goal is to show that this term does not contribute significantly to the time evolution of the system, and to quantify this contribution.

If we assume further that $g_0/\Omega_\text{m}\ll1$, we can approximate
\begin{equation}
\cos\bigl[2(\Omega_\text{m}+g_0\hat{A}_+)t^\prime\bigr]=\cos\bigl(2\Omega_\text{m}t^\prime\bigr)
\end{equation}
to first order in $g_0/\Omega_\text{m}$. At this order of approximation, we therefore have
\begin{align}
\hat{U}_\mathcal{DS}(t)=\hat{U}_\text{MO}(t)\,\overset{\leftarrow}{\mathcal{T}}\exp\biggl\{-\imath \int_0^t \rmd t^\prime \,\frac{g_0(t^\prime)}{2}\,\bigl[\cos(\Omega_\text{c}\,t^\prime)\,\hat{A}_++\sin(\Omega_\text{c}\,t^\prime)\,\hat{A}_-\bigr]\,\bigl[\cos\bigl(2\,\Omega_\text{m}\,t^\prime\bigr)\,\hat{B}_++\sin\bigl(2\,\Omega_\text{m}\,t^\prime\bigr)\,\hat{B}_-\bigr]\biggr\}.
\end{align}
to zeroth order in $1/\alpha$ and first order in $g_0/\Omega_\text{m}$. The full formula from which this expression was derived reads
\begin{multline}
\hat{U}_\mathcal{DS}(t)=\hat{U}_\text{MO}(t)\,\overset{\leftarrow}{\mathcal{T}}\exp\biggl(-\imath\,g_0\int_0^t \rmd t^\prime \bigl[2\,F_-(t^\prime)\,\cos(\Omega_\text{c}\,t^\prime)\,\hat{b}^\dag\hat{b}-2\,F_+(t^\prime)\,\sin(\Omega_\text{c}\,t^\prime)\,\hat{b}^\dag\hat{b}+\cos(\Omega_\text{c}\,t^\prime)\,\hat{A}_++\sin(\Omega_\text{c}\,t^\prime)\,\hat{A}_-\bigr]\\
\times\,e^{-4\,\imath\,F_\text{b}^{(2)}(t^\prime)}\,\hat{U}_-^\dag(t^\prime)\,\bigl\{\cos\bigl[2\,F_\text{b}(t^\prime)+2\,F_+\,\hat{A}_+(t^\prime)+4\,F_\text{b}^{(2)}(t^\prime)\,\hat{b}^\dag\hat{b}\bigr]\,\hat{B}_+\\
+\sin\bigl[2\,F_\text{b}(t^\prime)+2\,F_+\,\hat{A}_+(t^\prime)+4\,F_\text{b}^{(2)}(t^\prime)\,\hat{b}^\dag\hat{b}\bigr]\,\hat{B}_-\bigr\}\,\hat{U}_-(t^\prime)\biggr).
\end{multline}
From this formula we can see that the entire exponential in the time-ordered part is multiplied by $\frac{g_0}{\Omega_\text{m}}$ which is assumed to be significantly smaller than $1$. Therefore, this exponential contributes to at least first order. More importantly, from the reasoning below, we argue that the genuine second order corrections, those denoted by $\hat{h}_2$, will not contribute. Therefore, we can safely approximate all functions to their zeroth order, which means $F_{\pm}=F_\text{b}^{(2)}\approx0$ and $F_\text{b}\approx\eta$. All our results are valid for $\Omega_\text{m}\,t\ll1$.

\section{Error from neglecting correction in $\hat{U}_\mathcal{DS}(t)$}
First, introduce $\lvert\psi(t)\rangle:=\hat{U}(t)\lvert\psi(0)\rangle$ and $\lvert\chi(t)\rangle:=\hat{U}_0\lvert\psi(0)\rangle$ for some arbitrary initial state $\lvert\psi(0)\rangle$. We want to compute an approximate expression for the fidelity $\mathcal{F}(t):=\lvert\langle\chi(t)\vert\psi(t)\rangle\rvert^2$. We consider the rather generic case where 
\begin{align}
\label{eq:Appendix:Uexpansion}
\hat{U}(t)=\hat{U}_0(t)&\biggl(\mathds{1}-\frac{\imath}{\hbar}\,\int_0^t \rmd t^\prime\,\hat{h}_1\,\epsilon+\frac{1}{\hbar^2}\,\int_0^t \rmd t^\prime\,\hat{h}_1\,\int_0^{t^\prime}\rmd t^{\prime\prime}\,\hat{h}_1\,\epsilon^2-\frac{\imath}{\hbar^2}\,\int_0^t \rmd t^\prime\,\hat{h}_2\,\epsilon^2\biggr)+\mathcal{O}(\epsilon^3),
\end{align}
where the parameter $\epsilon$ is such that $\epsilon\ll1$, and where $\hat{h}_i$ ($i=1,2,3$) are generic hermitian operators. Notice, now, that there are two contributions to second order in $\epsilon$ in the above expression. The first comes from the square of the first-order contribution. The second is a genuine second-order contribution.

For convenience of presentation we rewrite \eref{Appendix:Uexpansion} as
\begin{align}
U(t)=&\hat{U}_0(t)\,\biggl(\mathds{1}-\frac{i}{\hbar}\,\hat{E}_1\,\epsilon-\frac{1}{\hbar^2}\,\overset{\leftarrow}{\hat{E}}_{11}\,\epsilon^2-\frac{i}{\hbar^2}\,\hat{E}_2\,\epsilon^2\biggr)+\mathcal{O}(\epsilon^3)\nonumber\\
U(t)^\dag=&\biggl(\mathds{1}+\frac{i}{\hbar}\,\hat{E}_1\,\epsilon-\frac{1}{\hbar^2}\,\overset{\rightarrow}{\hat{E}}_{11}\,\epsilon^2+\frac{i}{\hbar^2}\,\hat{E}_2\,\epsilon^2\biggr)\,\hat{U}_0(t)^\dag+\mathcal{O}(\epsilon^3),
\end{align}
where we have introduced
\begin{align}
\hat{E}_1(t)&:=\int_0^t\rmd t^\prime\hat{h}_1(t^\prime),\nonumber\\
\overset{\leftarrow}{\hat{E}}_{11}(t)&:=\int_0^tdt^\prime\int_0^{t^\prime}dt^{\prime\prime}\hat{h}_1(t^\prime)\hat{h}_1(t^{\prime\prime})\nonumber\\
\overset{\rightarrow}{\hat{E}}_{11}(t)&:=\int_0^tdt^\prime\int_0^{t^\prime}dt^{\prime\prime}\hat{h}_1(t^{\prime\prime})\hat{h}_1(t^\prime)\nonumber\\
\hat{E}_2(t)&:=\int_0^t\rmd t^\prime\hat{h}_2(t^\prime).
\end{align}
Proceeding, it is straightforward to see that, to second order in $\epsilon$,
\begin{align}
\label{eq:Appendix:FidelityIntermediate}
\mathcal{F}(t)&=\lvert\langle\chi(t)\vert\psi(t)\rangle\rvert^2\\
&=\lvert\langle\chi(t)\vert\hat{U}(t)\vert\psi(0)\rangle\rvert^2\\
&=\lvert\langle\chi(t)\vert\hat{U}_0(t)\vert\psi(0)\rangle\rvert^2-\frac{\imath}{\hbar}\langle\chi(t)\vert\hat{U}_0(t)\vert\psi(0)\rangle\langle\psi(0)\vert\hat{E}_1(t)\hat{U}_0^\dag(t)\vert\chi(t)\rangle\epsilon+\frac{\imath}{\hbar}\langle\psi(0)\vert\hat{U}_0^\dag(t)\vert\chi(t)\rangle\langle\chi(t)\vert\hat{U}_0(t)\hat{E}_1(t)\vert\psi(0)\rangle\epsilon\nonumber\\
&\qquad-\frac{\imath}{\hbar^2}\langle\chi(t)\vert\hat{U}_0(t)\vert\psi(0)\rangle\langle\psi(0)\vert\hat{E}_2(t)\hat{U}_0^\dag(t)\vert\chi(t)\rangle\epsilon+\frac{\imath}{\hbar^2}\langle\psi(0)\vert\hat{U}_0^\dag(t)\vert\chi(t)\rangle\langle\chi(t)\vert\hat{U}_0(t)\hat{E}_2(t)\vert\psi(0)\rangle\epsilon\nonumber\\
&\qquad-\frac{1}{\hbar^2}\langle\chi(t)\vert\hat{U}_0(t)\vert\psi(0)\rangle\langle\psi(0)\vert\overset{\rightarrow}{\hat{E}}_{11}(t)\hat{U}_0^\dag(t)\vert\chi(t)\rangle\epsilon^2-\frac{1}{\hbar^2}\langle\psi(0)\vert\hat{U}_0^\dag(t)\vert\chi(t)\rangle\langle\chi(t)\vert\hat{U}_0(t)\overset{\leftarrow}{\hat{E}}_{11}(t)\vert\psi(0)\rangle\epsilon^2\nonumber\\
&\qquad+\frac{1}{\hbar^2}\langle\chi(t)\vert\hat{U}_0(t)\hat{E}_1(t)\vert\psi(0)\rangle\langle\psi(0)\vert\hat{E}_1(t)\hat{U}^\dag_0(t)\vert\chi(t)\rangle\epsilon^2.
\end{align}
Let us recall, however, that we want to study the particular scenario where $\lvert\chi(t)\rangle=\hat{U}_0(t)\lvert\psi(0)\rangle$. Then, \eref{Appendix:FidelityIntermediate} simplifies dramatically to
\begin{equation}
\mathcal{F}(t)=1+\frac{1}{\hbar^2}\lvert\langle\psi(0)\vert\hat{E}_1(t)\vert\psi(0)\rangle\rvert^2\epsilon^2-\frac{1}{\hbar^2}\langle\psi(0)\vert\overset{\leftarrow}{\hat{E}}_{11}(t)\vert\psi(0)\rangle\epsilon^2-\frac{1}{\hbar^2}\langle\psi(0)\vert\overset{\rightarrow}{\hat{E}}_{11}(t)\vert\psi(0)\rangle\epsilon^2,
\end{equation}
which also guarantees formally that the fidelity $\mathcal{F}(t)$ satisfies $\mathcal{F}(t)\leq1$. 

To simplify this equation, notice that $\overset{\rightarrow}{\hat{E}}_{11}(t)+\overset{\leftarrow}{\hat{E}}_{11}(t)=\hat{E}^2_1(t)$. The proof of this statement is rather straightforward. Either it is clear from the unitarity definition of the time evolution and the fact that it must satisfy $U(t)\,U(t)^\dag$ at all orders, or
\begin{equation}
\frac{\rmd}{\rmd t}\biggl[\overset{\rightarrow}{\hat{E}}_{11}(t)+\overset{\leftarrow}{\hat{E}}_{11}(t)\biggr]=\hat{E}_1(t)\hat{h}_1+\hat{h}_1\hat{E}_1(t),
\end{equation}
and
\begin{equation}
\frac{\rmd}{\rmd t}\hat{E}^2_1(t)=\hat{E}_1(t)\hat{h}_1+\hat{h}_1\hat{E}_1(t),
\end{equation}
meaning that these two quantities are identical up to a constant. Since all these operators are equal to zero for $t=0$, then this constant must also be equal to zero. This proves the statement. Therefore we have
\begin{equation}\label{correction:formula:new}
\mathcal{F}(t)=1+\frac{1}{\hbar^2}\Bigl[\lvert\langle\psi(0)\vert\hat{E}_1(t)\vert\psi(0)\rangle\rvert^2-\langle\psi(0)\vert\hat{E}^2_1(t)\vert\psi(0)\rangle\Bigr]\epsilon^2,
\end{equation}
which is non-negative, always smaller than or equal to $1$, and manifestly real---as expected and required.

To proceed we define a dimensionless time variable $\eta=\Omega_\text{m}t$ (and, similarly, $\eta^\prime=\Omega_\text{m}t^\prime$). Given the form of our time evolution operator $\hat{U}(t)\to\hat{U}_\mathcal{DS}(t)$, and that $\epsilon=g_0/\Omega_\text{m}\ll1$, we have
\begin{multline}
\label{eq:Appendix:Ffull}
\mathcal{F}(\eta)=1-\biggl(\frac{g_0}{\Omega_m}\biggr)^2\Biggl(\langle\psi(0)\vert\Biggl\{\int_0^\eta\rmd\eta^\prime\Biggl[\hat{A}_+\cos\Biggl(\frac{\Omega_\text{c}}{\Omega_\text{m}}\eta^\prime\Biggr)+\hat{A}_-\sin\Biggl(\frac{\Omega_\text{c}}{\Omega_\text{m}}\eta^\prime\Biggr)\Biggr]\Bigl[\hat{B}_+\cos(2\eta^\prime)+\hat{B}_-\sin(2\eta^\prime)\Bigr]\Biggr\}^2\vert\psi(0)\rangle\\
-\Biggl\{\int_0^\eta\rmd\eta^\prime\langle\psi(0)\vert\Biggl[\hat{A}_+\cos\Biggl(\frac{\Omega_\text{c}}{\Omega_\text{m}}\eta^\prime\Biggr)+\hat{A}_-\sin\Biggl(\frac{\Omega_\text{c}}{\Omega_\text{m}}\eta^\prime\Biggr)\Biggr]\Bigl[\hat{B}_+\cos\bigl(2\eta^\prime\bigr)+\hat{B}_-\sin\bigl(2\eta^\prime\bigr)\Bigr]\vert\psi(0)\rangle\Biggr\}^2\Biggr),
\end{multline}
which is correct \textit{only} for $(g_0/\Omega_\text{m})\eta\ll1$.

Let us simplify \eref{Appendix:Ffull}. We start by computing $\hat{E}_1(\eta)$. We have
\begin{align}\label{final:partial:formula:explicit}
\hat{E}_1(\eta)=F_{++}(\eta)\hat{A}_+\hat{B}_++F_{--}(\eta)\hat{A}_-\hat{B}_-+F_{+-}(\eta)\hat{A}_+\hat{B}_-+F_{-+}(\eta)\hat{A}_-\hat{B}_+,
\end{align}
where we have introduced the functions
\begin{align}
F_{++}(\eta)&:=\frac{1}{2}\,\int_0^\eta d\eta^\prime\,\frac{g_0(\eta^\prime)}{\Omega_\text{m}}\,\cos\Biggl(\frac{\Omega_\text{c}}{\Omega_\text{m}}\,\eta^\prime\Biggr)\,\cos\bigl(2\,\eta^\prime\bigr)\nonumber,\\
F_{+-}(\eta)&:=\frac{1}{2}\,\int_0^\eta d\eta^\prime\,\frac{g_0(\eta^\prime)}{\Omega_\text{m}}\,\cos\Biggl(\frac{\Omega_\text{c}}{\Omega_\text{m}}\,\eta^\prime\Biggr)\,\sin\bigl(2\,\eta^\prime\bigr)\nonumber,\\
F_{-+}(\eta)&:=\frac{1}{2}\,\int_0^\eta d\eta^\prime\,\frac{g_0(\eta^\prime)}{\Omega_\text{m}}\,\sin\Biggl(\frac{\Omega_\text{c}}{\Omega_\text{m}}\,\eta^\prime\Biggr)\,\cos\bigl(2\,\eta^\prime\bigr)\nonumber,\ \text{and}\\
F_{--}(\eta)&:=\frac{1}{2}\,\int_0^\eta d\eta^\prime\,\frac{g_0(\eta^\prime)}{\Omega_\text{m}}\,\sin\Biggl(\frac{\Omega_\text{c}}{\Omega_\text{m}}\,\eta^\prime\Biggr)\,\sin\bigl(2\,\eta^\prime\bigr).
\end{align}
When $g_0$ is constant these can be calculated explicitly to yield
\begin{align}
F_{++}(\eta)&=\frac{\Omega_\text{m}}{2}\vastl[\frac{\sin\Bigl(\frac{\Omega_\text{c}-2\Omega_\text{m}}{\Omega_\text{m}}\eta\Bigr)}{\Omega_\text{c}-2\Omega_\text{m}}+\frac{\sin\Bigl(\frac{\Omega_\text{c}+2\Omega_\text{m}}{\Omega_\text{m}}\eta\Bigr)}{\Omega_\text{c}+2\Omega_\text{m}}\vastr],\\
F_{--}(\eta)&=\frac{\Omega_\text{m}}{2}\vastl[\frac{\sin\Bigl(\frac{\Omega_\text{c}-2\Omega_\text{m}}{\Omega_\text{m}}\eta\Bigr)}{\Omega_\text{c}-2\Omega_\text{m}}-\frac{\sin\Bigl(\frac{\Omega_\text{c}+2\Omega_\text{m}}{\Omega_\text{m}}\eta\Bigr)}{\Omega_\text{c}+2\Omega_\text{m}}\vastr],\\
F_{+-}(\eta)&=\frac{\Omega_\text{m}}{2}\vastl[\frac{\cos\Bigl(\frac{\Omega_\text{c}-2\Omega_\text{m}}{\Omega_\text{m}}\eta\Bigr)}{\Omega_\text{c}-2\Omega_\text{m}}-\frac{\cos\Bigl(\frac{\Omega_\text{c}+2\Omega_\text{m}}{\Omega_\text{m}}\eta\Bigr)}{\Omega_\text{c}+2\Omega_\text{m}}\vastr]-\frac{2\Omega_\text{m}^2}{\Omega_\text{c}^2-4\Omega_\text{m}^2},\ \text{and}\\
F_{-+}(\eta)&=\frac{\Omega_\text{m}}{2}\vastl[\frac{\cos\Bigl(\frac{\Omega_\text{c}-2\Omega_\text{m}}{\Omega_\text{m}}\eta\Bigr)}{\Omega_\text{c}-2\Omega_\text{m}}+\frac{\cos\Bigl(\frac{\Omega_\text{c}+2\Omega_\text{m}}{\Omega_\text{m}}\eta\Bigr)}{\Omega_\text{c}+2\Omega_\text{m}}\vastr]-\frac{\Omega_\text{c}\Omega_\text{m}}{\Omega_\text{c}^2-4\Omega_\text{m}^2}.
\end{align}
The second term in \eref{Appendix:Ffull} is more complicated and requires the evaluation of sixteen terms; its explicit expression is not illuminating. However, we note that, after some calculations, it is possible to see that it has the general form
\begin{align}
\mathcal{F}(\eta)=1-F_\text{uni}(\eta)-\biggl(\frac{g_0}{\Omega_m}\biggr)^2\langle\psi(0)\vert\hat{E}_\text{NO}(\eta)\vert\psi(0)\rangle,
\end{align}
where $\hat{E}_\text{NO}(\eta)$ is an operator made of \textit{normal-ordered} sestic, quartic, and quadratic combinations of the operators $\hat{a}^\dag$, $\hat{a}$, $\hat{b}^\dag$, and $\hat{b}$, while $F_\text{uni}(\eta)$ is a function of $\eta$ but independent of the initial state that can be obtained after some algebra. In particular, if we start with the vacuum for both modes $\hat{a}$ and $\hat{b}$, i.e., $\lvert\psi(0)\rangle=\lvert0\rangle$, we obtain the vacuum fidelity $\mathcal{F}_\text{vac}(\eta)$ that reads
\begin{equation}
\mathcal{F}_\text{vac}(\eta)=1-F_\text{uni}(\eta).
\end{equation}
Furthermore, it can be shown that
\begin{align}
F_\text{uni}(\eta)&=2\bigl[F_{++}(\eta)-F_{--}(\eta)\bigr]^2+2\bigl[F_{+-}(\eta)+F_{-+}(\eta)\bigr]^2\\
&=2\biggl(\frac{g_0}{\Omega_m}\biggr)^2\Biggl[\eta^2\sinc^2\Biggl(\frac{\Omega_\text{c}+2\Omega_\text{m}}{\Omega_\text{m}}\eta\Biggr)+\Biggl(\frac{\Omega_\text{c}-2\Omega_\text{m}}{2\Omega_\text{m}}\Biggr)^2\eta^4\sinc^4\Biggl(\frac{\Omega_\text{c}-2\Omega_\text{m}}{2\Omega_\text{m}}\eta\Biggr)\Biggr].
\end{align}
This expression is used, in terms of the time variable $t$, in the main text.
\twocolumngrid

\end{document}